\documentclass[aps,prl%
,twocolumn%
,showpacs%
]{revtex4}

\usepackage{graphicx}
\usepackage{times}

\begin{document}

\setlength{\textfloatsep}{1.2\parindent plus 2pt minus 2pt} 
\setlength{\arraycolsep}{1pt}
\newcommand{\be}{\begin{equation}}
\newcommand{\ee}{\end{equation}}
\newcommand{\bea}{\begin{eqnarray}}
\newcommand{\eea}{\end{eqnarray}}
\newcommand{\f}{\frac}
\newcommand{\p}{\partial}
\newcommand{\no}{\nonumber}
\newcommand{\kT}{k_{\rm B}T}
\newcommand{\e}{{\rm e}}
\newcommand{\dd}{{\rm d}}

\newcommand{\kav}{\left<k\right>}

\newcommand{\affila}{
 Biological Physics Research Group of HAS and Department\ of Biological Physics, E\"otv\"os University,
 P\'azm\'any P.\ stny.\ 1A, H-1117 Budapest, Hungary
}
\newcommand{\affilb}{
}

\title{Statistical mechanics of topological phase transitions in networks}
\author{Gergely Palla}
 \affiliation{\affila}
\author{Imre Der\'enyi}
 \affiliation{\affila}
\author{Ill\'es Farkas}
 \affiliation{\affila}
\author{Tam\'as Vicsek}
 \affiliation{\affila}

\date[]{\protect\today}

\begin{abstract}

We provide a phenomenological theory for topological transitions
 in restructuring  networks. In this statistical mechanical approach 
 energy is assigned to the different network topologies and temperature is
 used as a quantity referring to the level of noise during the
 rewiring of the edges. The associated microscopic dynamics satisfies the 
 detailed balance condition and is equivalent to a lattice gas model on the
 edge-dual graph of a fully  connected network. In our studies -- based on
 an exact enumeration method, Monte-Carlo simulations, and theoretical
 considerations -- we find a rich variety of topological
 phase transitions when the temperature is varied.
These transitions signal singular changes in the essential features of the 
global structure of the network. Depending on the energy function chosen, the observed transitions can be best
 monitored using the
 order parameters $\Phi_s=s_{\rm max}/M$, {\it i.e.}, the size of the
 largest connected component divided by the number of edges,
 or $\Phi_k=k_{\rm max}/M$, the
 largest degree in the network divided by the number of edges.
 If, for example the energy is chosen to be
 $E=-s_{\rm max}$, the observed transition is analogous to
 the percolation
 phase transition of random graphs.
For this choice of the energy, the phase-diagram in 
 the $[\kav,T]$ plane is constructed. Single vertex energies of the form
 $E=\sum_i f(k_i)$, where $k_i$ is the degree of vertex $i$, are also
 studied. Depending on the form of $f(k_i)$, first order and continuous 
phase transitions can be observed.
 In case of $f(k_i)=-(k_i+\alpha)\ln(k_i)$, the transition is continuous,
 and at the critical temperature scale-free graphs can be recovered. Finally,
 by abruptly decreasing the temperature, non-equilibrium processes 
({\it e.g.}, nucleation,
 growth of particular topological phases) can also be interpreted by the present approach.

\end{abstract}



\maketitle

\section{I. Introduction}

In recent years, the analysis of the network structure of interactions 
has become a popular and fruitful method used in the study of
 complex systems. Whenever many similar objects  in mutual
 interactions are encountered, these objects can be represented 
 as nodes and the interactions as links between the nodes, defining a network.
 The world-wide-web, the science citation index, and biochemical
 reaction pathways in living cells are all good examples of complex systems 
 widely modeled with networks, and the set of further phenomena where 
the network approach can be used is even more diverse.
 In most cases, the overall structure of
 networks reflect the
 characteristic properties of the original systems, and 
 enable one to sort seemingly
 very different systems into a few major classes of stochastic graphs
 \cite{b-a-revmod,dorog-mendes-book}.
 These developments have greatly
advanced the potential to interpret the fundamental common features of
such diverse systems as social groups, technological, biological and
other networks. The effects of both the restructuring
\cite{watts2} and
the growth \cite{b-a-science} of the associated graphs have been
considered, leading to a number of exciting discoveries about the laws
concerning their diameter, clustering and degree distribution.
Real networks typically exhibit both aspects (growth and rearrangement) one
 of which is usually dominating the dynamics. Here we concentrate on the 
evolution of graphs due to restructuring, but shall briefly discuss the growth
 regime as well.

Various interesting effects observed in networks can be interpreted 
using analogies with well understood phenomena studied in statistical physics.
 As a classical example, we mention the percolation 
 phase transition in the Erd\"os-R\'enyi (ER) random graph model
 \cite{ErdosRenyi,bollobas},
 which occurs by varying the average degree, $\kav$, of the
 vertices around $\kav=1$. For
$\kav<1$ the graph falls apart into small pieces, on the other hand
for $\kav\geq 1$ a giant connected component emerges (in addition to the 
finite components).
 Another subtle example is the mapping of a growing network model
onto an equilibrium Bose gas \cite{bianconi}.
 For the latter model, under certain conditions 
 a single node is allowed to collect a finite fraction of all edges,
 corresponding to
 a highly populated ground level and sparsely populated higher energies
 seen in Bose-Einstein condensation. 

When connecting the graph theoretical aspects of networks
 to statistical physics, one can step further from the analogies
 by directly defining  {\em statistical ensembles for graphs}.
 The use of a statistical mechanical formalism  for the changes
in graphs being in an equilibrium-like state is expected to provide a
significantly deeper insight into the processes taking place in systems
being in a saturated state and, as such, dominated by the fluctuating
rearrangements of links between their units.

As an example, let us take a given number of units interacting in a
``noisy'' environment. These units can be people, firms, genes, {\it etc}. The
probability for establishing a new or ceasing an existing
interaction between two units depends on both the noise and
the advantage gained (or lost) when adopting the new
configuration. 
In this picture, a global transition in the connectivity properties can occur
 as a function
of the level of noise. For instance, if the conditions are such
that the interactions between the partners become more ``conservative''
(safer choices are more highly valued), then -- as we show
later -- a transition from a less ordered to a more ordered network
 configuration can take place. In particular, it has been argued
 \cite{stark-vedres} that depending on the level of certain types of 
  uncertainties (expected fluctuations) business networks reorganize from a
 star-like topology to a system of more cohesive, highly clustered ties.

There are several possible ways to define the statistical ensemble of 
networks. In \cite{burda,burda2},
 the members of the ensemble were identified
 by the Feynman diagrams of a field theory in zero
 dimensions (called ``minifield''), and the weights of the graphs
 were given by the corresponding
amplitudes calculated using the standard Feynman rules. The ensemble obtained
this way was characterized by the fractal and spectral dimensions, and the 
 dependence of the topology of the graphs on these two parameters was
 discussed. The authors argued that in the parameter plane of two 
 parameters related to the fractal and spectral dimension, the region of
 generic graphs and the region
 of crumpled graphs are  separated by a line; and on this separating
 line scale-free
 networks appear.
An alternative definition for the partition function was proposed by 
  Berg and L\"assig  in \cite{berg}, resulting in a simpler formalism,
 analogous to the statistical mechanics of classical Hamiltonian systems.
They introduced a Hamiltonian for networks, and also a 
 parameter $\beta$ playing the role of inverse temperature. The weights
 of different graphs in the partition function were obtained from 
these two quantities as in classical statistical mechanics.
 These studies showed that Hamiltonians beyond the single vertex form 
(where terms depending on connectivities between the vertices also appear) 
 lead to correlations between the  vertices for large $\beta$.
 A similar model leading
 to interesting results was presented in \cite{manna}, where the Hamiltonian
 depended on the ratios of the degrees of neighboring vertices, and
 the dynamics favored disassortative mixing and high clustering. The system
 organized itself into three phases depending on one parameter: the exponential, scale-free and hub-leaves states were produced, respectively. However, the non-uniform
 selection of links at the rewiring in this model makes it impossible
 to satisfy the detailed balance condition.

In this paper we  analyze the reorganization of networks from the
 point of view of {\em topological phase transitions}, {\it i.e.}, 
transitions in the graph structure as a function of {\em temperature},
 the quantity representing the level of noise during the
restructuring process of the network. 
 For clarity we note that
our studies concern a class of phenomena that are clearly different
from  the phase transitions investigated
 by applying models of statistical mechanics originally defined
 on regular lattices to an underlying (static) random network structure 
\cite{rodriguez,miguel,goltsev}, or the phase transitions observed in
 growing networks \cite{bianconi,kertesz}, or quasi-static networks 
\cite{manna2}.
Topological phase transitions are accompanied by singularities in the
 thermodynamic functions derived from the partition function of the 
 statistical graph ensemble and can be characterized by a drastic change in an 
 appropriate order parameter.
 Our statistical ensemble (similarly to the one presented
  in Ref.\ \cite{berg}), is defined by introducing 
  an {\em energy} that 
 accounts for the advantage or loss during the rearrangement. In our
 description, this energy may depend on either
  global properties of the network, or single vertex degrees as well. 

The use of Hamiltonian formalism also provides a general frame for
 the optimization of network structure (for examples of network
 optimization problems see \cite{optimalize,soft-opt}). To find the optimal
 configuration for a given task,  the system has to
 be cooled, using an appropriate energy function. 

This article is a direct extension of our previous work \cite{sajat};
covering more details, results, and new approaches.
 The paper is organized as
 follows: in Sec. II we define the canonical ensemble of the networks
 together with the partition function
 and other essential thermodynamic quantities. In Sec. III we discuss the 
numerical methods used to study the phase transitions. In Sec. IV we 
present the phase transitions obtained for energies that depend
 on global properties, and Sec. V is devoted to two interesting cases of
 single vertex energies.  In Sec. VI we discuss briefly  the grand canonical
 ensemble of networks and we conclude in Sec. VII.

\section{II. Statistical mechanics of networks}

We shall consider a set, $\{ g_a \}$, of 
{\em undirected graphs}, 
containing $N$ nodes and $M$ links.
Each graph $g_a$ can be 
represented by the {\em adjacency matrix}
 $A_{ij}^a$, where $A_{ij}^a=1$ if vertices $i$ and
 $j$ are connected and 
it is zero otherwise.
In a heat bath at temperature $T$, the {\em canonical ensemble} of these
 graphs (in analogy with that proposed in \cite{berg}), 
can be defined by the partition function 
\be
Z(T)=\sum_{ \{ g_a \} } \e^{-E_a / T},
\label{Z}
\ee
where $E_a$ is the {\em energy} assigned to the different
 configurations.

 The restructuring processes of the network 
can be interpreted
 via the following physical picture: The
basic event of rearrangement is the reallocation of a randomly
selected edge (link) to a new position either by ``diffusion'' (keeping
one end of the edge fixed and connecting the other one with a new node)
or by removing the given edge and connecting two randomly selected
nodes. Then, the energy difference
$\Delta E_{ab}=E_b-E_a$
between the original $g_a$ and the new $g_b$ configurations is
calculated and the reallocation is carried out
following the Metropolis algorithm \cite{metropolis}.
 If the energy of the new graph
 is lower than that of the original one, the reallocation is accepted; if 
 the new energy is higher, the reallocation is accepted
 only with
 probability $\e^{-\Delta E_{ab}/T}$. This way, in the $T\to\infty$ limit 
the dynamics converges to a totally random rewiring
process, and thus, the classical ER random graphs are recovered.
 On the other hand, at low temperatures the
topologies with lowest energy occur with enhanced probability.
 The resulting dynamics, 
by construction, satisfies the detailed balance condition \cite{metropolis}.

\begin{figure}[t!]
\centerline{\includegraphics[angle=0,width=0.9\columnwidth]{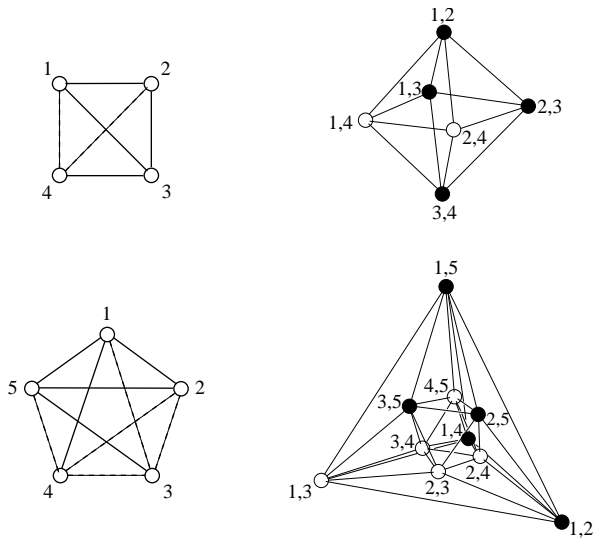}}
\caption{ Two simple examples of graphs (left hand side) and the corresponding 
edge-dual graphs (right hand side).
 The full spheres in the edge-dual graphs represent occupied sites,
 corresponding to existing bonds in the original graph
 (drawn with solid lines),
 and the hollow spheres are the empty sites,
 corresponding to absent bonds (represented by dashed lines) in the original 
 network. The rewiring of an edge in the original graphs is equivalent to
 the displacement of the corresponding particle on the edge-dual graph.
}
\label{edge_dual}
\end{figure}
 This network rearrangement is formally equivalent to a {\em
Kawasaki type lattice gas} dynamics with conserved number of particles
moving on a special lattice, which is the edge-dual graph of the fully
connected network
\cite{bollobas,ramezanpour}.
The sites of this lattice are the possible
$N(N-1)/2$ connections between the vertices, and the particles wandering
on the sites are the $M$ edges, as shown in Fig. \ref{edge_dual}.

The partition
function (\ref{Z}) contains many terms corresponding to {\em topologically
equivalent} graphs: these graphs can be simply transformed into one another
 by an  adequate permutation of the indexing. 
 Since we consider energies $E_{\alpha}$ that depend only on the topology
 $t_{\alpha}$,
 it is natural to rewrite
 the partition
function in a form where the summation runs through all possible topologies:
\bea
Z(T)=\sum_{\lbrace t_{\alpha}\rbrace}{\cal N}_{\alpha} \e^{-E_{\alpha}/T}.
\label{topologicalZ}
\eea
Here we introduced ${\cal N}_{\alpha}$
 to count the number of configurations belonging to topology $t_{\alpha}$. 
Expression (\ref{topologicalZ}) can be rewritten as
\bea
Z(T)&=&\sum_{\lbrace t_{\alpha} \rbrace }\e^{-E_{\alpha}/T+
\ln({\cal N}_{\alpha})}
=\sum_{\lbrace  t_{\alpha}\rbrace}\e^{-F_{\alpha}/T},\\
F_{\alpha}&=&E_{\alpha}-TS_{\alpha},\\
S_{\alpha}&=&\ln({\cal N}_{\alpha}),\label{S_a}
\eea
where $F_{\alpha}$ is the free energy  and 
$S_{\alpha}$ is the entropy of the topology $t_{\alpha}$. 

We are interested in the possible singularities in the thermodynamic
 functions derived from the partition function above, since they, if there are
 any, correspond to phase transitions in the topology of the associated
 networks.
These transitions can be 
best monitored by introducing a suitable 
{\it order parameter}. As we are primarily
interested in the transitions between dispersed and compact states, a
natural choice can be either
$\Phi=\Phi_s=s_{\rm max}/M$, the number of edges of the largest
connected component of the graph
$s_{\rm max}$ normalized by the total number of edges $M$, or
$\Phi=\Phi_k=k_{\rm max}/M$, the highest degree in the graph
$k_{\rm max}$ divided by $M$. We also introduce the
corresponding conditional free energy $F(\Phi,T)$ via
\be
\e^{-F(\Phi,T) / T}=Z(\Phi,T)=\sum_{ \{ g_a \}_\Phi } \e^{-E_a / T},
\label{FPhi}
\ee
where $\{ g_a \}_\Phi$ is a subset of $\{ g_a \}$, consisting of all the
graphs with order parameter $\Phi$. A phase transition, where 
 a rapid change occurs in the order parameter from $\Phi=0$ towards higher 
 values, is also 
accompanied by a shift
of the minimum of the conditional free energy $F(\Phi,T)$.
 A sudden change in the position of the global
minimum signals a discontinuous (first order) phase transition, whereas a
gradual shift indicates either a crossover or a continuous
phase transition. 

In the next section we briefly discuss the numerical methods used to
 study  topological phase transitions.

\section{III. Numerical methods}
\noindent
{\em III. A)  Exact enumeration method}\\

The numerical results shown in this paper were obtained by two alternative
methods. Motivated by the success of a similar approach used in
 random-walks, percolation, and polymers related problems 
\cite{guttmann,havlin,havlin-book,stanley},
 for small systems we evaluated the partition function together with the
 probability of every individual state via an
 {\em exact enumeration method}. In this approach, first all possible
 connected configurations with a given number of edges are generated
 successively
 up to  $M$: the graphs with $m+1$ edges are constructed from the
 graphs with $m$ edges either by linking a new vertex to one of
 the old vertices, or by linking two previously unconnected vertices. 
Next, all possible configurations containing $M$ edges are obtained from
 the combination of smaller connected graphs, with sizes up to $M$. 
Finally, ${\cal N}_{\alpha}$ is calculated for each topology $t_{\alpha}$ 
by counting the number
of possible permutations chosen to label the vertices
in the state. (For more details and a simple example, see appendix A).
 The probability $p_{\alpha}$, of a 
topology $t_{\alpha}$ then can be obtained from 
\bea
p_{\alpha}=\f{{\cal N}_{\alpha}\e^{-E_{\alpha}/T}}{Z}.
\eea
Once the set of possible
states with appropriate probabilities has been constructed, one can evaluate 
the expectation value of any ${\cal Q}$ thermodynamic quantity  using
\bea
\left<{\cal Q}\right>=\sum_{t_{\alpha}} {\cal Q}(t_{\alpha})p_{\alpha}.
\eea
The advantage of this technique, beside
 producing exact results, is that the set of ${\cal N}_a$ has to be
 calculated only once, independently of the energy functions considered,
 in contrast to 
Monte-Carlo simulations, where the
 simulation has to be restarted from the beginning every time we introduce
 a new type of energy.
 This method is limited by the rapid growth of the number of topologies with
$M$.
 For networks
 of size seen in the real world ($M>10^2$) the realization of this method
 is clearly unfeasible. \\

\noindent 
{\em III. B) Monte-Carlo simulations} \\

The lattice gas model defined on the edge-dual graph of the fully connected network relaxes slowly,
because interactions are dense \footnote{
To judge the density of interactions, 
recall that among the sites of the underlying lattice 
-- {\it i.e.}, the edges of the complete graph with $N$ vertices --
there are only first and second neighbors.
One lattice site (one edge of the complete graph) 
has $2(N-2)$ first neighbors and 
$N^2/2-5N/2+3$ second neighbors.}
and the energy
minima are sometimes localized in hardly accessible
parts of the phase space of the system. A good example is the
transition from a classical random graph 
-- stable at high temperatures --
to a star, which is stable at low temperatures. 
The simplest Monte-Carlo rewiring
simulation (discussed in Sec. II.) tries to move a randomly chosen edge to a randomly chosen new
location. However, this method is very inefficient, if one would like
to simulate the condensation of edges into a star in a large system.

There are several simulation tools 
that can help to achieve faster convergence. 
We have used the so-called parallel tempering 
(also called exchange Monte-Carlo) method in several cases
\cite{parallel-tempering}. Except for 
first-order transitions, this algorithm can be used well 
to measure the transition at an acceptable speed and high
precision. 

The algorithm can be viewed as an improved version of simulated
annealing. 
Several replicas of the system are simulated simultaneously,
and each of them is connected to a separate heat bath.
 A replica in a hot heat bath will explore the "large-scale"
structure of phase space, and the motion of 
a replica in a cold heat bath will be restricted to a small part of
phase space, where it will explore the deep but narrow energy wells.

In the exchange Monte-Carlo method, after a given number of 
conventional update steps within each replica,
exchange steps are made. Two replicas (with neighboring
temperatures) are chosen at random, and a Monte-Carlo-type decision is
made whether the two replicas should be exchanged, {\it i.e.}, their
temperatures should be swapped. 
With Metropolis dynamics,
if the product of the energy difference between the two replicas ($\Delta E$)
and the difference of inverse temperatures ($\Delta\beta$) is positive, then 
this exchange is accepted, otherwise it is
accepted only with probability $\e^{\Delta \beta \,\Delta E}$.
%
That is, a replica with a high energy and a
low inverse temperature ({\it i.e.}, high temperature) will be more likely
to remain in its own heat bath, whereas a replica with a high energy
and a high inverse temperature ({\it i.e.}, low temperature) will be likely
to be "put" into a heat bath with a higher temperature.

 We shall now move on to review some of energy functions,
that lead to phase transitions, when the temperature is changed from
 zero to infinity at constant $\kav$. 
Since at $T=\infty$ the entropically favorable
 graphs dominate, a minimum
 requirement for the energy function is that the configurations with the
 lowest energy must also have low entropy. We divide the investigated
 energy functions into two categories. In the first category we put
the energies that depend on component sizes in the graph,
the other group contains 
the single-vertex energies.

%
%
%

\section{IV. Cluster energies}

As mentioned in the introduction, the classical random graph model
 (corresponding to $T\to\infty$) exhibits a phase transition
 when the average degree of the vertices, $\kav=2M/N$, 
is varied around $\kav=1$.
 For $\kav<1$, the network consists of small, disconnected clusters, on the
other hand, for $\kav\geq 1$ a giant connected component emerges in the graph
 collecting a finite portion of the edges. Near
the critical point the size of the giant component scales as
$(\kav-1)M$.

Based on the lattice gas analogy we expect that if $\kav<1$, then for a
suitable choice of the energy (one that rewards clustering) a similar
dispersed-compact phase transition occurs at a finite temperature
$T(\kav)$. Such a transition can be best monitored by the order
parameter
$\Phi_s=s_{\rm max}/M$ (the number of edges in the largest connected 
   component $s_{\rm max}$ divided by the total number of edges),
 often used in graph theory \cite{newman-phase}.

The most obvious energy satisfying the above requirement is a
monotonically decreasing function $E=f(s_{\rm max})$. In this case the
 energy is independent 
of the distribution of the size of smaller clusters, or of the structural
 details of  the largest cluster: only 
the size of the largest cluster matters.
The entropic part of the conditional free energy in this case can
 be estimated by counting the number of configurations at given 
$s_{\rm max}$. The number of different connected configurations of
 size $s_{\rm max}$ can be estimated as $s_{\rm max}^{s_{\rm max}}$ to
 leading order \cite{bollobas2}, (for an intuitive derivation see 
 appendix B). 
This term has to be multiplied
 by the number of possible selections of these $s_{\rm max}$
vertices out of $N$, which is simply $N$ over
 $s_{\rm max}$.
The $M-s_{\rm max}$ left-out edges can be placed anywhere between the 
$N-s_{\rm max}$ remaining vertices, with the restriction that they cannot
 form clusters larger than $s_{\rm max}$.
 Since we consider $s_{\rm max}=\Phi_sM$ to be an extensive quantity
and the typical size of the largest component of these left-out edges scales
 slower than $M$, this constraint can be neglected. Hence, the contribution
from the left-out edges can be well estimated by an
 $(N-s_{\rm max})^2/2$ over $(M-s_{\rm max})$ factor.
 If we combine these
factors together, then in the thermodynamic limit
 (when $N,M\rightarrow\infty$, $\kav=\mbox{const.}$) we can write 
\bea
{\cal N}_{\rm clus}\approx (\Phi_sM)^{\Phi_sM}
{N\choose \Phi_sM}
{(N-\Phi_sM)^2/2 \choose M-\Phi_sM}.
\eea
Since the energy of the system is a function of the order parameter $\Phi_s$ 
itself, the conditional free energy $F(\Phi_s,T)$ can be expressed as
\bea
\e^{-F(\Phi_s,T)/T}={\cal N}_{\rm clus}\e^{-f(\Phi_s)/T}=
\e^{-[f(\Phi_s)-T\ln {\cal N}_{\rm clus}]/T}.
\eea
By  using 
Stirling's formula [$l!\approx (l/e)^l\sqrt{2\pi l}$] to approximate the factorials and neglecting terms 
of ${\cal O}(\ln N)$, for $\ln {N}_{\rm clus}$ we get
\bea
\ln{\cal N}_{\rm clus}&\approx&\mbox{const.}
+\left[1+\ln \f{2M}{N}-\f{2M}{N}\right]\Phi_sM\no \\ 
& &+\left[\f{3M}{2N}-\f{1}{2}-\f{M^2}{N^2}\right]\Phi_s^2M.
\label{S_clus}
\eea
By replacing $2M/N$ with $\kav$ in the expression above,
 the resulting conditional
 free energy can be expressed as
\bea
F(\Phi_s,T) &\approx&
 f(\Phi_s M)
 +MT\Biggl\lbrace\left[\kav-1-\ln(\kav)\right]\Phi_s\Biggr.
\no\\
 &&\Biggl.+\left[\kav^2-3\kav+2\right]\f{\Phi_s^2}{4}\Biggr\rbrace
\label{Fs}.
\eea

\begin{figure}[t!]
\centerline{\includegraphics[angle=-90,width=0.9\columnwidth]{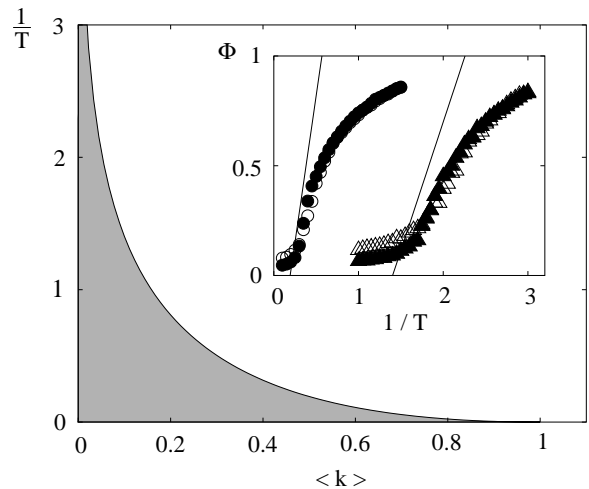}}
\caption{
The phase diagram and the order parameter for the $E=-s_{\rm max}$ energy.
Main panel: The white and shaded areas correspond to the ordered phase
(containing a giant component) and the disordered phase, respectively, 
as given by Eq. (\ref{border}).
Inset: The order parameter $\Phi=\Phi_s=s_{\rm max}/M$
obtained from Monte-Carlo simulations
as a function of the inverse
temperature for $\langle k\rangle=0.1$ (triangles) and 
$\langle k\rangle=0.5$ (circles).
Each data point is an ensemble average of $10$ runs, time averaged
between $t=100N$ and $500N$ Monte-Carlo steps. The open and closed
symbols represent $N=500$ and $1,000$ vertices, respectively.
The critical exponent, in agreement with the
analytical approximations (solid lines),
was found to be $1$.
}
\label{fig_s}
\end{figure}

The simplest choice for an energy function that depends only on the 
size of the largest component is
\bea
f(s_{\rm max})=-s_{\rm max}=-\Phi_sM.
\eea
In this case it can be clearly seen from
Eq.\ (\ref{Fs}) that as long as
\bea
T>T_{\rm c}(\kav)=\f{1}{\kav-1-\ln(\kav)},\label{border}
\eea
the free energy has a minimum at
$\Phi_s=\Phi_s^*(T)=0$, {\it i.e.}, the configuration is
dispersed (see main panel of Fig.\ \ref{fig_s}). 
When the temperature drops below
$T_{\rm c}(\kav)$, the minimum moves away from $\Phi_s=0$ and a giant
component appears. Near the critical temperature $T_{\rm c}(\kav)$, the
order parameter at the minimum of the free energy can be estimated from
Eq.\ (\ref{Fs}) as
\bea
\Phi_s^*(T)=2\f{T^{-1}-T^{-1}_{\rm c}(\kav)}{\kav^2-3\kav+2},
\eea
indicating that we are dealing with a {\it continuous topological phase
transition} (see inset of Fig.\ \ref{fig_s}).

\begin{figure}[t!]
\centerline{\includegraphics[angle=0,width=0.9\columnwidth]{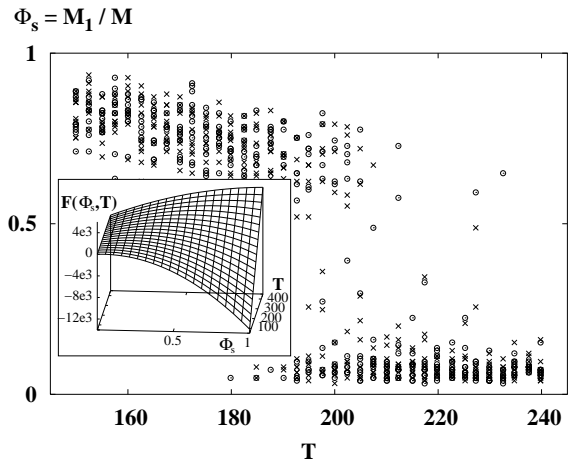}}
\caption{
If the energy of the graph is $E=-s_{\rm max}^2$, then the
order parameter, $\Phi_s=M_1/M$, shows a first order transition.
($M_1$ is the number of edges in the largest component of the graph.)
Each point gives the value of $\Phi_s$ averaged between 
$t=490N$ and $t=500N$ Monte-Carlo
steps in a graph started at $t=0$ 
from an Erd\H os-R\'enyi random
graph (o) or a star ($ \times $). The simulated graph had
$N=500$ vertices and $M=125$ edges.
 The inset shows the behavior of the analytical free energy obtained from
 (\ref{Fs}) using the same parameters. The temperature interval in which the
 two minima (at $\Phi_s=0$, corresponding to the connected state and at
 $\Phi_s=1$, corresponding to the dispersed state) coexist is 
 fully compatible with the numerical findings for the transition regime.
}
\label{s1sqr}
\end{figure}

Topological phase transitions of first order are also expected to occur for
other forms of cluster energies. For example, when 
$f(s_{\rm max})$ starts with a zero 
(or positive) slope.
 In such a case, the behavior of the conditional free energy 
 at very high and very low temperatures is similar to the previous case:
 when $T\rightarrow\infty$, the energy term $f(s_{\rm max})$ can 
 be neglected in (\ref{Fs}), and the entropic term has
 a minimum at $\Phi_s=0$, which is the dispersed state. In contrast,
 when  $T\rightarrow0$, only the energy term remains,
 resulting a minimum in $F(\Phi_s,T)$ 
 at $\Phi_s=1$, the compact state. However, there is also an intermediate
 temperature range, where 
$F(\Phi_s,T)$ given by (\ref{Fs}) starts as an increasing function 
at $\Phi_s=0$ (since the linear term in the entropy dominates for small
 $\Phi_s$),
 then reaches its maximum somewhere in the $[0,1]$ interval and
continues as a decaying function from that point on 
(since the higher order decaying terms in the energy overcome the 
increasing terms at larger $\Phi_s$). As a consequence, 
 the conditional free energy has two competing minima in the $[0,1]$ interval,
a meta-stable and a globally stable. The coexistence of stable and meta-stable
 minima at the transition between the phases is a characteristic of 
 first order phase transitions. 
A simple function of this type is $f(s_{\rm max})=-s_{\rm max}^2$.
For this choice of the energy, our numerical results are 
shown in Fig.\ref{s1sqr}.
 The  hysteresis appearing
between cooling and heating supports the theoretical considerations 
 about the coexisting minima that indicate a first order phase transition,
 summarized in the inset of Fig.\ref{s1sqr}. The analytical conditional 
 free energy gained by substituting $f(s_{\rm max})=-s_{\rm max}^2$ into 
(\ref{Fs}) has  a single minimum, when $T<80$ and when $T>430$, in former case
 at the dispersed state ($\Phi_s=0$), in latter case at the  connected state
 ($\Phi_s=1$). At intermediate temperatures  these two minima coexist, 
 predicting a  first order phase transition somewhere in the middle part of 
this temperature interval. The transition regime $170< T <270$ observed in 
the MC simulation is compatible with the 
 analytical result, since one does not expect the simulation to reveal
 the meta-stable state beside its dominant stable counterpart in
  case of a very significant difference between the depths of the according
  minima in the free energy.

We have also investigated the case of
 $f(s_{\rm max})=-s_{\rm max}\ln(s_{\rm max})$, both analytically and numerically. Similarly
 to the previous case, it can be shown that there is a temperature regime,
 where the conditional free energy as a function of $\Phi_s$ has two
 competing minima, hence the transition is of first order. 

\begin{figure}[t!]
\centerline{\includegraphics[angle=-90,width=0.9\columnwidth]{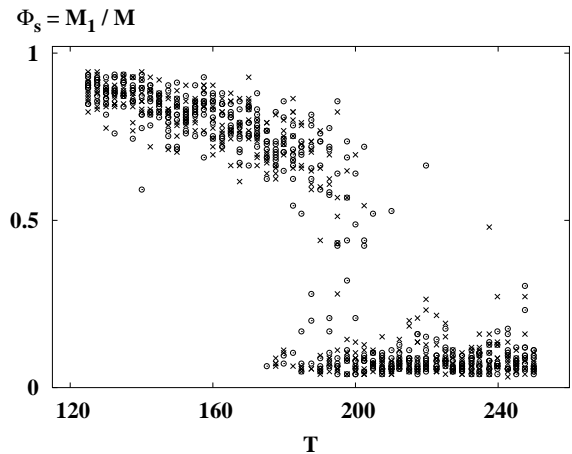}}
\caption{
The order parameter, $\Phi_s=M_1/M$,
for the $E=-\sum_{i=1}^{N_c} s_i^2$ energy.
($M_1$ is the number of edges in the largest component and 
$N_c$ is the number of components in a graph.)
Each point shows the value of $\Phi_s$ after $t=200N$ Monte-Carlo
steps in a graph started at $t=0$ 
from an Erd\H os-R\'enyi random
graph (o) or a star ($ \times $). The simulated graph had
$N=500$ vertices and $M=125$ edges.
Observe that for intermediate temperatures there are two distant
groups of states with high stability
(at $\Phi_s\approx 0.6-0.9$ and $\Phi_s\approx 0-0.15$),
and $\Phi_s$ values in the region between these two
rarely occur, {\it i.e.}, a first order transition was found.
}
\label{fig_sisqr}
\end{figure}

Since the energy $E=f(s_{\rm max})$ depends on a global quantity (the
size of the largest connected component) it might be also reasonable
to define the
energy of the graph as $E=\sum_j f(s_j)$, where the summation goes over
each component and $s_j$ denotes the number of edges in the $j$th
one. The total number of edges, $M=\sum_j s_j$, is conserved
by the dynamics, hence $f(s_j)$ must decrease faster than
linear to promote compactification.
 When a single giant component (containing the
majority of the edges) emerges, its energy
$f(s_{\rm max})$ dominates the energy of the entire graph, and,
as a good approximation, the above analysis for $E=f(s_{\rm max})$ can
be repeated, leading to first order and continuous phase transitions. 
For the case of $E=-\sum_{i=1}^{N_c} s_i^2$, our numerical results
 are presented in Fig.\ref{fig_sisqr}, showing a first order phase transition.

\begin{figure}[t!]
\centerline{\includegraphics[angle=-90,width=1\columnwidth]{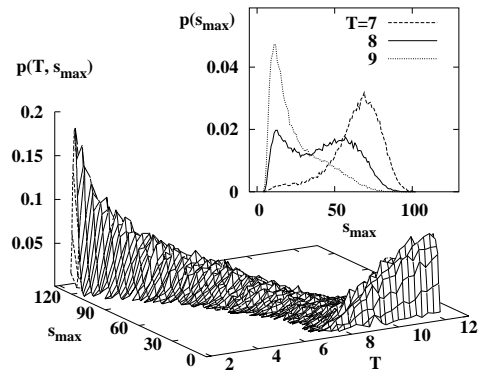}}
\caption{
Distribution of the size of the largest graph component, 
$s_{\rm max}$,
if the $E=-\sum_{i=1}^{N}s_i\ln(s_i)$ energy is used. 
Main panel. 
At low temperatures the distribution of the largest component 
has one maximum at large values of $s_{\rm max}$, 
this is the ordered phase.
At high temperatures, the largest component is small: the graph
is disordered. 
Inset. 
Using a higher resolution, one can observe
that the transition from the low-temperature peak ($T=7$) 
to the high-temperature peak ($T=9$) happens via 
a bimodal distribution
($T=8$, indicated by a solid line). 
This indicates that 
the conditional free energy of the system has two 
competing minima at the
intermediate temperature, and 
the transition is of first order.
The graphs used for the simulations had 
$N=500$ vertices and $M=125$ edges. 
Averages were taken for $10$ (main panel) or $200$ (inset)
simulation runs between 
simulation times of $t=200N$ and $t=400N$ Monte-Carlo steps
using time steps of $t=N$.
}
\label{silnsiP}
\end{figure}

For the $E=-\sum_{i=1}^{N}s_i\ln(s_i)$ energy -- similarly to the
case of the $E=-s_{\rm max}\ln(s_{\rm max})$ energy -- we found a
first-order transition between the ordered phase 
(present at low temperatures) and
the disordered phase (high temperatures).
Results are shown in Fig. \ref{silnsiP}.


\section{V. Single vertex energy functions}

Next we turn to another important class of the energy functions, where the
energies are assigned to the vertices rather than to the connected
components of the graph:
\bea
E=\sum_{i=1}^{N}f(k_i), \label{f-sforma}
\eea
where $k_i$ denotes the degree (number of neighbors) of vertex $i$.
This energy is consistent with a dynamics in which the change of the
degree of a vertex depends only on the structure of the graph in its
vicinity. The fitness
of an individual vertex depends on its
connectivity. The most suitable order parameter for this class of
graph energy is
$\Phi=\Phi_k=k_{\rm max}/M$.
Again, due to the conservation of the number of edges,
$M=\sum_i k_i$, the single vertex energy $f(k_i)$ should decrease
faster than $-k_i$, if aggregation is to be favored.

We introduce an alternative form for single vertex energies:
\bea
E=\sum_{i=1}^{N}\sum_{i'}g(k_{i'}), \label{g-sforma}
\eea
where $i'$ runs over all vertices that are neighbors of vertex $i$.
In this interpretation, the fitness of an individual vertex
 depends on the connectivities of its neighbors, and 
 vertex $i$ collects an energy $g(k_{i'})$ from each of its neighbors. 
These neighbors in turn will all collect $g(k_i)$ from vertex $i$, 
therefore the total contribution to the energy from vertex $i$ is 
$k_i g(k_i)$.
Thus, by using
\bea
f(k_i)=k_i g(k_i),
\eea
 the two alternative forms of the single vertex energy, (\ref{f-sforma})
 and (\ref{g-sforma}), become equivalent.\\

\noindent 
{\em V. A) The energy $E=-\sum k_i^2$ : mapping to the Ising-model} \\

A natural choice for the energy of single vertex type is the following.
 Assign the negative 
energy $-J$ to all pairs of edges that share a common vertex at one end.
 The total energy of a given configuration is then
\bea
E=-\f{J}{2}\sum_{i=1}^Nk_i(k_i-1)=-\f{J}{2}\sum_{i=1}^Nk_i^2+\f{1}{2}JM,
\label{E_k2}
\eea
corresponding to $f(k_i)= -(J/2)k_i^2$,
(or equivalently, to $g(k_i)=-(J/2)k_i$). The constant term in
 (\ref{E_k2}) does not play any role in the dynamics, hence it can be omitted.
 This form of the energy is in full analogy with the usual
definition of the  
energy
\bea
E=-J\sum_{<\alpha,\beta>} n_{\alpha} n_{\beta}
\eea
of a lattice gas on the edge-dual graph of the fully connected network
with nearest neighbor attraction. The summation here runs over all
adjacent pairs of lattice sites (corresponding to possible edges between the
 vertices of the original graph),
 and $n_{\alpha}=1$ if site $\alpha$ is occupied
and $0$ otherwise. When this energy is applied to lattices,
 we recover the standard lattice gas model of nucleation of
vapors. The negative energy unit $-J$ associated with a pair of edges
sharing a vertex in the original graph is equivalent to the binding
energy between the corresponding occupied nearest neighbor sites on the
edge-dual graph. By measuring the energies (and temperature) in units
of $J$ we can set $J=1$, without loosing generality. Thus, from now on
$J$ will be omitted.

 The lattice gas representation 
can be further transformed to an Ising-model-representation
by introducing the $s_{\alpha}\in[-1,1]$ spin-like variables connected to 
$n_{\alpha}$ as $n_{\alpha}=(1+s_{\alpha})/2$.
The energy with the help of the spins is expressed as
\bea
E&=&-\f{1}{4}\sum_{<\alpha,\beta>}s_{\alpha}s_{\beta}-
\f{1}{2}\sum_{\alpha=1}^{N(N-1)/2} s_{\alpha}\no \\& &
-\f{1}{8}N(N-1)(N-2),
 \label{EofIsing}
\eea
 since the total number of lattice sites equals $N(N-1)/2$, and the number of 
adjacent pairs of lattice sites is $N(N-1)(N-2)/2$.
This is similar to a ferromagnetic Ising-model in an external magnetic field.
 If the number of occupied sites in the lattice gas picture equals
 the number of unoccupied sites, the contribution from the external magnetic
 field vanishes in 
 the Ising-model picture. However, in the thermodynamic limit, where
 $N\rightarrow\infty,M\rightarrow\infty,\kav=2M/N=$const., this condition
 cannot be fulfilled, since  the total number of sites scales as $N^2$,
 whereas the number of particles scales as $M$ with the system size. 

For the particular form of $f(k_i)$ chosen, the topology with the lowest
 overall energy is
a ``star'' (for simplicity, we consider $M<N$),
 where all the $M$ edges are connected to single node.
The form of the conditional
 free energy in this case can be 
estimated as follows.
 In the $\Phi_k>1/2$ regime, where the system contains a star of size
 larger than $M/2$, the energy of this star dominates 
 the rest of the graph.
  Therefore,
 in the  thermodynamic limit, we neglect this latter contribution to the
 total energy and approximate the energy of the graph by that of the
 largest star.
Now we estimate the number of possible configurations for a given
 value of $\Phi_k$.
 In case of a star with $K=\Phi_kM$ arms, the central vertex 
can be chosen from $N$ different vertices.
 Once this is fixed, $K$ edges have to be
distributed among the $N-1$ possible links between the other vertices and 
the central one, yielding a factor of $N-1$ over $K$. The rest of the 
edges that are not part of the star can be placed anywhere between
 the $N-1$ (non-central) vertices, 
contributing a factor of $(N-1)(N-2)/2$ over $(M-K)$.
\bea
{\cal N}_{\rm star}(K) &\approx& N { N-1 \choose K}{(N-1)(N-2)/2 \choose M-K}
\approx \nonumber \\
& & N
{N\choose K}{N^2/2 \choose M-K}
\label{N_star}.
\eea
Again, we use 
Stirling's formula to approximate the factorials, and neglect
terms of the order ${\cal O}(\ln N)$ yielding
\bea
\ln{\cal N}_{\rm star}&\approx& M\Biggl[\f{N}{M}\ln\f{N}{M}+\f{N^2}{2M}\ln\f{N^2}{2M}
\Biggr.\no \\
& &
-\left(\f{N}{M}-\Phi_k\right)\ln\left(\f{N}{M}-\Phi_k\right)\no \\
& &-\Phi_k\ln\Phi_k-(1-\Phi_k)\ln(1-\Phi_k)\no \\
& &\Biggl.-\left(\f{N^2}{2M}-1+\Phi_k\right)\ln\left(\f{N^2}{2M}-1
+\Phi_k\right)\Biggr].
\eea
In the thermodynamic limit, to leading order we receive 
\bea
\ln {\cal N}_{\rm star}(x)\approx-\Phi_kM\ln(N),\label{S_str(x)}
\eea
where the $\Phi_k$ independent terms were dropped.
The resulting conditional free energy is expressed as:
\bea
F(\Phi_k,T)\approx 
 f(\Phi_k M)
 +\Phi_kMT\ln(N)\label{F_star}.
\eea
In the present case, with the $f(k_i)=-k_i^2$ energy, (\ref{F_star}) 
can be written as
\bea 
F(\Phi_k,T)\approx M\left[-\Phi_k^2M+\Phi_kT\ln(N)\right].
\label{F_star2}
\eea
Note that this approximation would be valid even for $\Phi_k<1/2$, if the
energy of the graph was simply defined as $E=f(k_{\rm max})$. 

\begin{figure}[t!]
\centerline{\includegraphics[angle=-90,width=0.9\columnwidth]{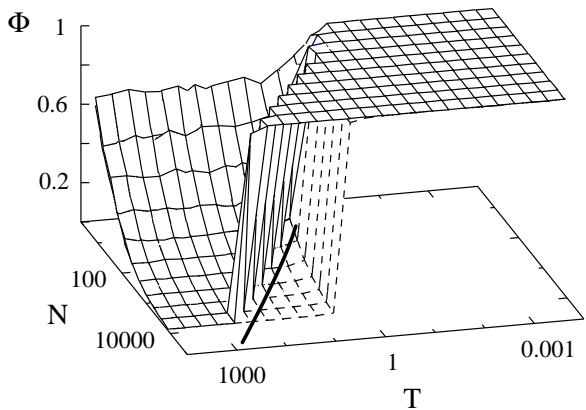}}
\caption{
The order parameter $\Phi=\Phi_k=k_{\rm max}/M$ as a function of
the temperature and the system size for $E=\sum_i-k_i^2/2$ and
$\langle k\rangle=0.5$.
The simulations were started either from a star 
(corresponding to $T=0$, solid line)
or a classical random graph 
($T=\infty$, dashed line). 
Each data point represents a single run, time averaged between
$t=100N$ and $200N$ Monte-Carlo steps. 
The thick solid line shows the analytically calculated
spinodal $T_1=M/\ln(N)$.}
\label{MC_k2result}
\end{figure}

The parabola given by Eq.\
(\ref{F_star2}) has a maximum at $\Phi_k=T/M\ln(N)$. When $T\to 0$, this
maximum also shifts towards zero and $F(\Phi_k,T)$ becomes a descending
parabola on the $[0,1]$ interval. This means that the minimum of the
free energy is at $\Phi_k=1$, the star configuration. In contrast, when
the temperature goes above the $T_1=M/\ln(N)$ spinodal point
(thick solid line in Fig.\ \ref{MC_k2result}),
the maximum moves out of the $[0,1]$
interval and the free energy becomes an ascending parabola, resulting
in a minimum at a low value of $\Phi_k$ (corresponding to an ER random
graph). However, this value cannot be deduced from Eq.\
(\ref{F_star2}), because it is a valid approximation only for
$\Phi_k>1/2$.
For intermediate temperatures the maximum of the parabola separates the
two extreme topologies: the dispersed random graph and the star). One
of these two extreme states is meta-stable and the other one is
stable.
 Due to the limited validity of Eq.\ (\ref{F_star2}), the
stability of these configurations can be studied only for
temperatures where the maximum of the parabola is well inside the $[1/2,1]$
 interval. 
\begin{figure}[t!]
\centerline{\includegraphics[angle=0,width=0.9\columnwidth]{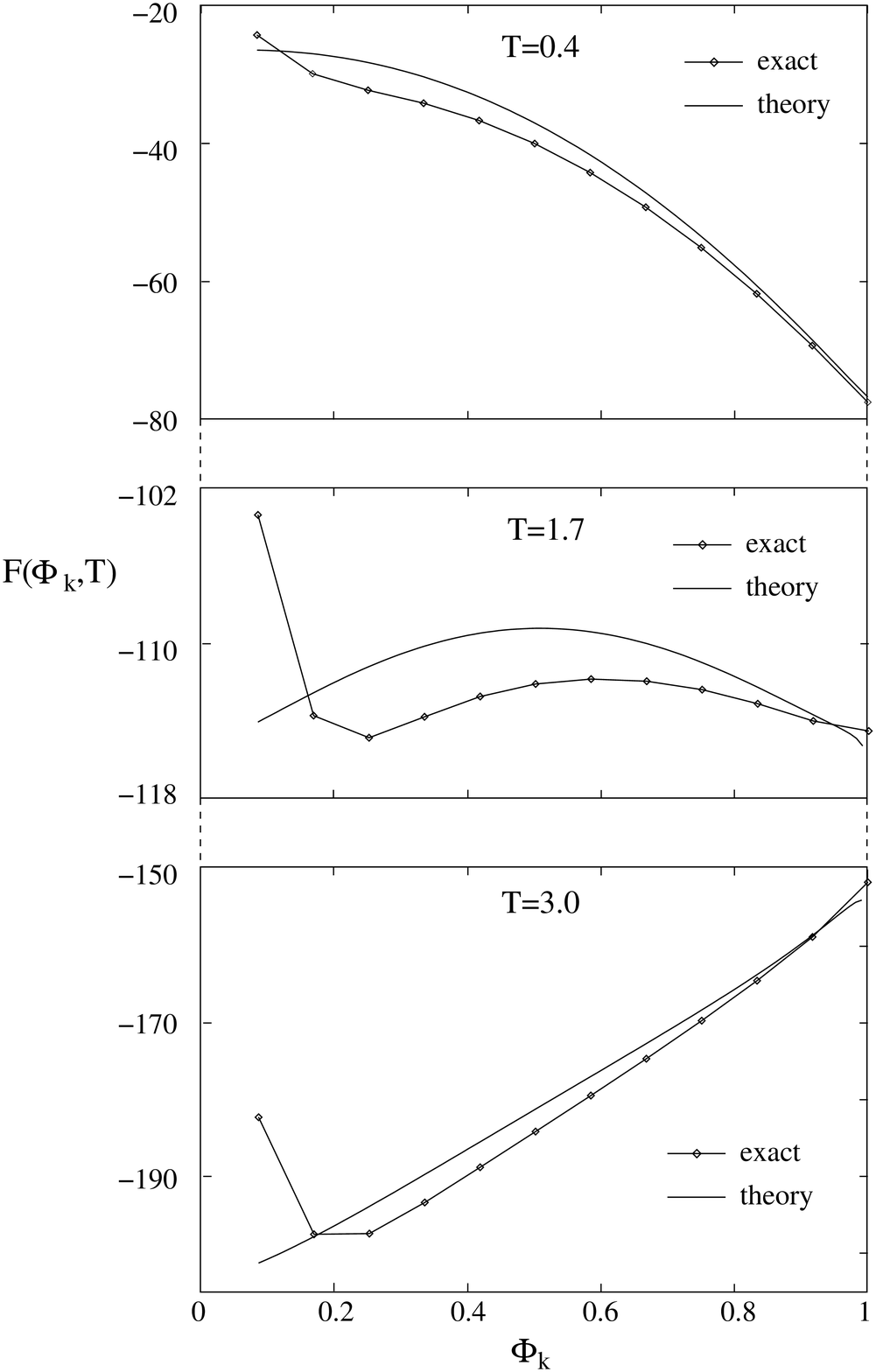}}
\caption{ The picture of the conditional free energy at three different
 temperatures for the $f(k_i)=-k_i^2$ energy, obtained from the
 exact enumeration method plotted together with the prediction of 
 our simple theoretical analysis for $M=12,N=48$. 
At low temperatures $F(\Phi_k,T)$ is a
 descending function on the $[0,1]$ with a minimum at $\Phi_k=1$, the
 star configuration (top figure), on the other hand for high temperatures,
 it becomes  
ascending for most part, with a minimum at low $\Phi_k$, the dispersed states
(bottom picture).
 There is an intermediate temperature regime in between,
 where the maximum of $F(\Phi_k,T)$ separates two competing minima 
(middle figure), hence this phase transition is of first order.} 
\label{ex_enum_fig}
\end{figure}
\begin{figure}[t!]
\centerline{\includegraphics[angle=0,width=0.9\columnwidth]{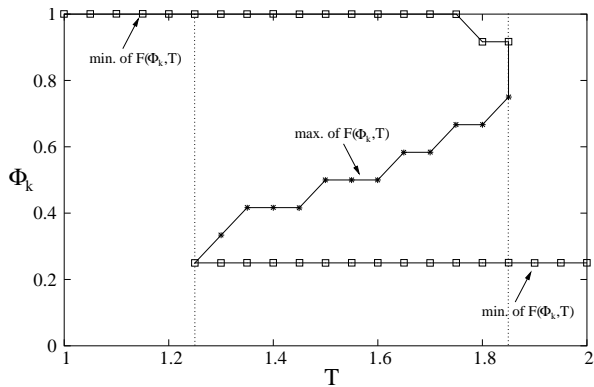}}
\caption{The spinodal curve obtained from the exact enumeration method with
 $E=-\sum_i k_i^2$, for $M=12,N=48$. At low and high temperatures,
 the conditional free energy $F(\Phi_k,T)$ has a single minimum (plotted with 
squares). At intermediate temperatures (in between the two dotted
 lines) there are two competing minima. In this latter temperature regime,
 the spinodal curve is obtained by plotting the maximum of $F(\Phi_k,T)$
 (represented by stars), besides the two minima.}
\label{spinoda}
\end{figure}

The scenario of the transition from a dispersed state to the star configuration
(see above) indicates that it is a { \em first order phase transition}.
 This is well supported by the results of both the exact enumeration method
 and Monte-Carlo  simulations. For small systems, the conditional free energy
 was evaluated via the exact enumeration method for various temperatures,
and was found to be in qualitative agreement with the prediction of the
 theoretical analysis, as demonstrated in Fig. \ref{ex_enum_fig}. The
three different temperature regimes described in the previous paragraph
 can be recognized in the behavior of the exact $F(\Phi_k,T)$ as well.
 Furthermore, in the intermediate temperature regime, where the conditional
 free energy has two competing minima, the spinodal curve can also be
 constructed as is shown in Fig. \ref{spinoda}. 
For large enough systems, in MC simulations a sudden change of the order
parameter between zero and one can be observed as shown in
Fig. \ref{MC_k2result}. The hysteresis appearing
between cooling and heating is consistent with a first order
transition.\\

\noindent 
{\em V. B) The energy $E=-\sum k_i\ln k_i$ : continuous phase-transition} \\

Another application-motivated  choice for the single vertex energy is
$f(k_i)=-k_i\ln(k_i)$, or equivalently, $g(k_i)=-\ln(k_i)$,
inspired, in part, by the logarithmic law of sensation. It is the
logarithm of the degree of a vertex that its neighbors can sense and
benefit from. In this case the configuration with the lowest energy is a
fully connected subgraph [or almost fully connected if $M$ cannot be
expressed as $n(n-1)/2$]. On the other hand, the star configuration is
 also quite favorable, since the energy of both the maximal possible star and
 of the  maximal possible fully connected subgraph scales as $-M\ln M$ to 
leading order.
 Amongst the sub-dominant terms in the energy,
 there is a difference in the order of $\sqrt{M}\ln \sqrt{M}$
 between the two, in favor of the fully connected subgraph. 
 As before,
we choose the order parameter to be
 $\Phi=\Phi_k=k_{\rm max}/M$, since this can easily
distinguish between these two configurations: $k_{\rm max}\approx \sqrt{2M}$
 for a fully connected subgraph counting $M$ edges, and
$k_{\rm max}\approx M$ for a star. 

\begin{figure}[t!] 
\centerline{\includegraphics[angle=-90,width=0.9\columnwidth]{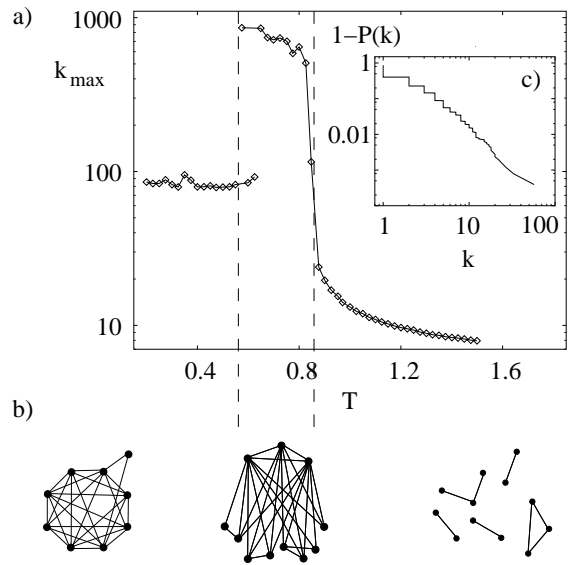}}
\caption{
Phases of the graph when the energy is $E=-\sum_i k_i\ln(k_i)$.
(a) The largest degree $k_{\rm max}$
for $N=10,224$ vertices and $M=2,556$ edges.
Each data point represents a single run, time averaged between
$t=5,000N$ and $20,000N$ MC steps.
The data points are connected to guide the eye.
There is a sharp, continuous transition near
$T=0.85$ and a first-order transition (with a hysteresis) 
around $T=0.5-0.6$.
(b) The three different plateaus in (a)
correspond to distinct topological phases:
$k_{\rm max}={\cal O}(1)$ to the classical random graph,
$k_{\rm max}={\cal O}(M)$ to the star phase
(a small number of stars sharing most of their neighbors)
and 
$k_{\rm max}={\cal O}(\sqrt{M})$ to the fully connected subgraph.
(c) The (cumulative) degree distribution at $T=0.84$ and $t=600N$
follows a power law.
This shows that the
degree distribution decays
as a power-law with the exponent $\gamma\approx 3$.
}
\label{MC_kmax}  
\end{figure}

Our MC simulations demonstrate (Fig.\ \ref{MC_kmax}) that as we cool
down the system, first the edges of the dispersed random graph assemble to
 form a configuration with a few large stars (sharing most of their neighbors),
and then at lower temperatures the graph is rearranged into an almost fully
connected subgraph. This is consistent with the fact that beside the slight
energetical disadvantage, the star configuration is entropically more favorable
when compared to the fully connected subgraph; therefore the 
latter configuration can take over only at very low temperatures.
 The hysteresis near the few large star {\it vs.} fully connected subgraph
  transition suggests
that it is a {\em first order phase transition}.
 On the other hand, the transition between the dispersed state and
 the few large stars 
 is accompanied by a singularity in the heat
 capacity (also seen with the exact enumeration method), and no hysteresis is 
observed, indicating that it is a {\em continuous phase transition}. 

For $\Phi_k>1/2$ Eq.\ (\ref{F_star}) can be used again as a good
approximation for the free energy of the graph, since the compact
 cluster arising from the dispersed state is rather star like. By plugging
$f(\Phi_k M)=-(\Phi_k M)\ln(\Phi_k M)$
into that expression, we get
\bea
F(\Phi_k,T)\approx M(T-1)\ln(N)\Phi_k
\eea
to leading order, which is linear in $\Phi_k$. In agreement with our
observations above, this formula predicts that for $T<1$ the star is a
stable configuration ($\Phi_k$=1 is a minimum of the free energy), and
for $T>1$ it becomes unstable. The transition at $T=T_{\rm c}=1$ is
thus step-like with no hysteresis, indicating a continuous phase
transition with an infinitely large critical exponent. We assume that the
 observed deviation of $T_{\rm c}$ from 1 in the MC simulations is
 a finite size effect.\\

\noindent
{\em V.C) Relation to growth with preferential attachment}\\

A remarkable feature of the MC dynamics is that in case
 of the energy $f(k_i)=-k_i\ln k_i$, by crossing
$T_{\rm c}$ from above, a scale-free graph (with a degree distribution $\sim k^{-\gamma}$ with $\gamma\simeq 3$) appears at some point of the
evolution of the graph from the random configuration towards the star.
This supports the notion that scale-free graphs are temporary 
(dynamical) configurations, not typical in equilibrium distributions.
The MC dynamics is governed by the change of the energy associated with
the reallocation of an edge. Estimating the energy change of a vertex by
the derivative of the single vertex energy $f(k_i)=-k_i\ln(k_i)$, we
get $\Delta E=1-\ln(k_i)$. Plugging this into the Boltzmann factor,
$\exp[-\Delta E/T]$, at $T=T_{\rm c}=1$ we get a
quantity proportional to $k_i$ for the acceptation/rejection ratio of a
randomly selected move. Since the preferential attachment in the
Barab\'asi-Albert model
\cite{b-a-science}
is proportional to $k_i$, {\it it is  natural that our dynamics
also produces scale-free graphs}.

Another interesting aspect of the $f(k_i)=-k_i\ln k_i$ energy is that
 the configurations in the two compact phases resemble
 the two major graph topologies obtained in 
Ref. \cite{optimalize}, by optimizing
the network for local search with congestion. Our intermediate phase with
a few large central hubs sharing neighbors is similar to
 the optimal topology for a small number of parallel searches, whereas
 the low-temperature configuration, the fully connected subgraph
 resembles the homogeneous topology optimal for a large number of 
parallel searches. However, an important difference between the two problems
 is that in our case a vertex is allowed to lose all of its
 connections under the restructuring process. The two ``similar-to-optimum''
 configurations appear as a natural consequence of the underlying dynamics.
 This observation suggests a 
 potential application of the presented theory:
tackling problems related to graph topology optimization by
simulated annealing techniques.\\

\noindent
{\em V. D) Topology-dependent non-extensiveness of the energy}\\

Both types of the single vertex energy functions discussed in the present
 section lead to compact configurations at low temperatures, for
 which the most highly connected vertices possess macroscopic numbers of edges.
 As a consequence, {\em the energy of the system scales differently
 with system size at high and low temperatures},
and diverges differently as $N\rightarrow\infty$. At high temperature,
 the system consists of many small unlinked clusters of about the same size,
 therefore a change in the total system size affects only the 
 number of the clusters, and the energy scales as $N$. On the other hand,
 when $f(k_i)=-k_i\ln(k_i)$, at low temperatures the energy of the star
 and the fully connected subgraph scales as $N\ln(N)$; in case of
 $f(k_i)=-k_i^2$, the energy of the star scales as $N^2$.
 Thus (unlike, {\it i.e.}, in the mean-field Ising model), there
is {\it no way to choose an appropriate coupling constant that could render
the energy extensive in all topological states simultaneously}.

Nevertheless, the dispersed state (having an extensive graph energy) can
equally be studied in the grand canonical ensemble.

\section{VI. The grand canonical ensemble}

In the grand canonical ensemble, the degree
distribution can be expressed as \cite{berg}
\bea
P_k=C\f{\e^{-\beta f(k)-\mu k}}{k!}.
\label{p_nagykan}
\eea
where C is a normalization factor and the chemical potential $\mu$ is
adjusted to give the correct $\kav$. For $f(k)=-k\ln(k)$, using
Stirling's formula, the distribution takes the form
\bea
P_k=C\f{\e^{-(\mu-1)k}}{\sqrt{2\pi k}}k^{(1/T-1)k}.
\eea
When $T>1$, this has a tail, which decays faster than exponential,
consequently, each vertex has a small degree. For $T<1$, on the other
hand, the tail becomes divergent, signaling a phase transition at
$T=T_{\rm c}=1$. Note however that in the $T<1$ temperature range,
due to the non-extensive contribution of the diverging degrees, the
ensembles are not equivalent, and the grand canonical description
loses its validity.

At the critical temperature, the grand canonical description might
still be valid. Choosing a more general single vertex energy,
$f(k_i)=-(k_i-\alpha)\ln(k_i)$, and setting $\kav$ such that $\mu=1$,
the degree distribution acquires a power law tail
($P_k\sim k^{-(\alpha+1/2)}$) and the network becomes scale-free at
this temperature.
We have to stress though that the scale-free network at $T_{\rm c}$ is
not general: for $\mu>1$ the tail decays exponentially, and for $\mu<1$
the tail diverges.

\section{VII. Summary }

We studied the restructuring in networks using a canonical
ensemble, where temperature corresponds to the level of noise in
real systems and the energy associated with the different
configurations accounts for the advantage gained or lost during
the rewiring of the edges. 
We found that for various types of energies, 
first order and continuous phase transitions may appear when 
 changing temperatures.
In case of the $E=-s_{\rm max}$ energy, if $\kav<1$, a dispersed-loose
 phase transition occurs at a finite temperature, equivalent to the
 percolation phase transition of classical random graphs when 
 $\kav$ is varied around $\kav=1$. We obtained a simple
 expression for the  $T_c(\kav)$ critical line separating the
 two phases in the $[\kav,T]$ plane from a theoretical analysis
 of the conditional free energy. For other forms of the energy depending
 on the size of the largest cluster we found 
first order phase transitions. We also studied the effects of 
different single vertex energies, namely the $E=-\sum_ik_i^2$ and
 $E=-\sum_ik_i\ln(k_i)$  cases. The network in the former case exhibits
 a first order phase transition from a dispersed state to a star-like state,
where nearly all edges are linked to a single vertex. With the
 $-\sum_ik_i\ln(k_i)$ energy, the dispersed state transforms into 
a compact one
 with a few large stars via a continuous phase
 transition. In the critical point, scale-free networks can be recovered. 
 At lower temperatures another transition occurs (this time of
 first order), where the configuration is turned into a fully
 connected subgraph. 

Although in this paper we assumed that $\kav\leq1$, this is not a
necessary requirement, when the energy is assigned to individual vertices.
For large average degree ($\kav>2$) the only difference is that one vertex
cannot collect all the edges, and thus, several stars appear in the
``star'' configuration. Further interesting directions in the
context of the above study include the investigation of additional
relevant forms for the energy [{\it e.g.}, $E=(k-n)^2$ with $n>1.5$] and the
joint effects of restructuring and growth.

\section{VIII. Acknowledgments}
The authors are grateful to G\'abor Tusn\'ady for many valuable
discussions.
This research has been supported by 
the Hungarian Scientific Research Fund
under grant No: OTKA 034995. 
I. F. acknowledges a scholarship from the
Communication Networks Laboratory at ELTE.

\vspace{-5pt}                           

\section{Appendix A}

 In the exact enumeration method, as mentioned in sec.III.A, the first
 step is to generate all connected graphs with $m+1$ edges from the
 connected graphs with $m$ edges, either by connecting a new vertex to the 
 core or by introducing a new link. In order to avoid double counting, every 
 new graph obtained this way is compared one by one to all already 
 revealed topologies using the following algorithm. Two graphs of 
 identical topology have identical degree distribution also, therefore 
 this property is checked first. In case of perfect match, the vertices 
 in both graphs are labeled in such a way, that a given index belongs 
 to vertices with equal number of links in the two graphs. Next, for each 
 index in one graph, the set of the neighbors indices is compared to 
 its equivalent index set in the other graph. If not all sets are identical, 
 then the labels in one of the graphs  have to be permuted until perfect 
 match between the neighboring relations is reached. (Obviously, labels are
 interchanged between vertices of same degree only). If the perfect match
 in the neighboring relations cannot be achieved for any permutation of 
 the indices, the two graphs are of different topology. 
     
 When a new topology is obtained, the corresponding combinatorial factors
 can be generated in a similar manner, by counting the number of permutations
 of the indices in the graph that lead to the same neighboring relations
 (same neighboring index sets) as the original indexing.

As a simple example, we demonstrate the evaluation of ${\cal N}_{\alpha}$ for 
 all states in case of  $M=3,N\geq6$.
 The construction of the connected graphs, and
 the possible topologies are shown below:\\
\centerline{\includegraphics[angle=0,width=0.9\columnwidth]{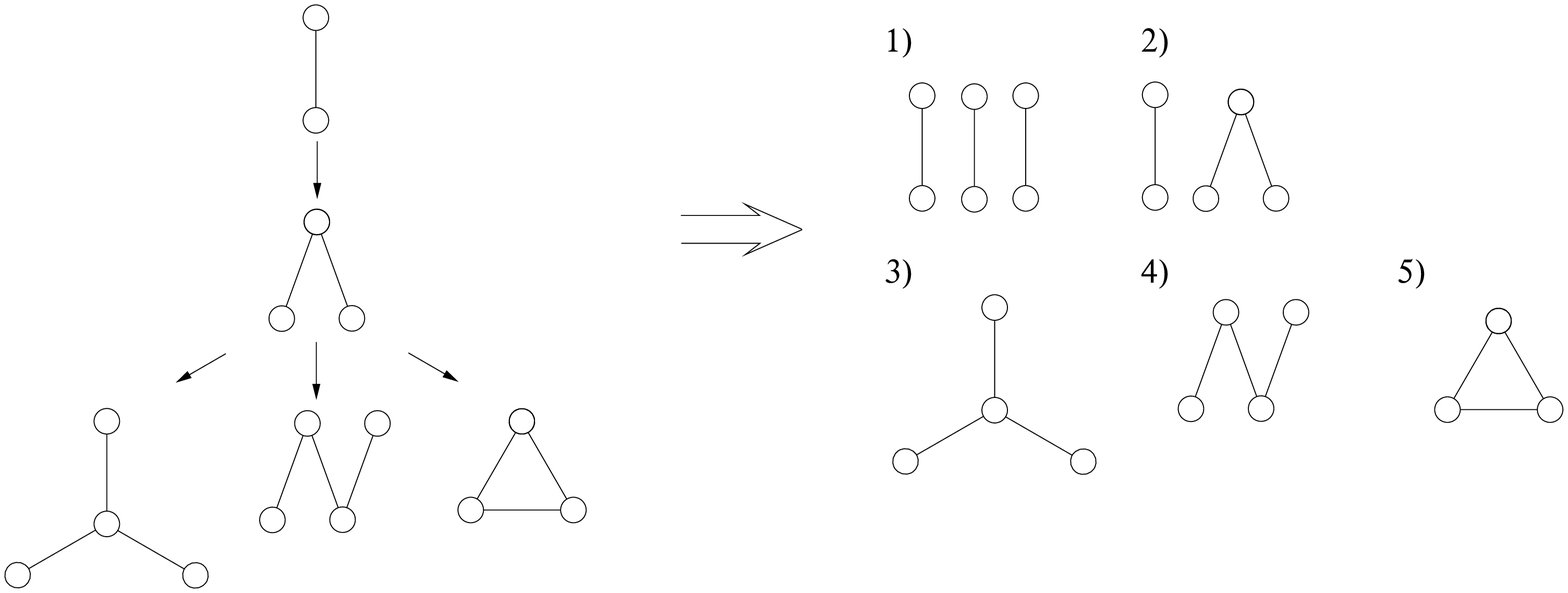}}

In case of a topology that does not possess any symmetries,
 ${\cal N}$ is simply 
$N!/(N-N_t)!$, where $N_t$ is the number of vertices included in the 
topology.
 In general this initial ${\cal N}$ has to be further
 divided by the number of those permutations of the indices of
 the vertices that leave the topology unchanged. Therefore, if
 the topology contains $n$ 
 identical subgraphs (like in case of state $\alpha=1$ above, 
where the topology
 is built up from three identical subgraphs) the initial value of ${\cal N}$
 has to be divided by $n!$. Furthermore, if any subgraph in the topology
 remains unchanged for $l$ permutations of the indices within 
itself, ${\cal N}$ has to be divided by $l$. In the example above,
 for the states $\alpha=1$ and $\alpha=2$, for all subgraphs, $l=2$,
 in case of the states $\alpha=3$ and $\alpha=5$, $l=3!$,
 and for the state $\alpha=4$, $l=2$.

Altogether, in the chosen case,
the ${\cal N}$ of the five possible states can be expressed as
\bea
& &{\cal N}_{1}=\f{N!}{(N-6)!2^33!}, \;\;
{\cal N}_{2}=\f{N!}{(N-5)!2^2},\;\;\no\\
& &{\cal N}_{3}=\f{N!}{(N-4)!3!}, \;\;
{\cal N}_{4}=\f{N!}{(N-4)!2},\;\;
{\cal N}_{5}=\f{N!}{(N-3)!3!}.\no
\eea

To provide a simple example of an application, we show the
 first few most probable states in case of $E=-\sum k_i\ln k_i$ at $T=0.65$ :\\
\\
\centerline{\includegraphics[angle=0,width=0.9\columnwidth]{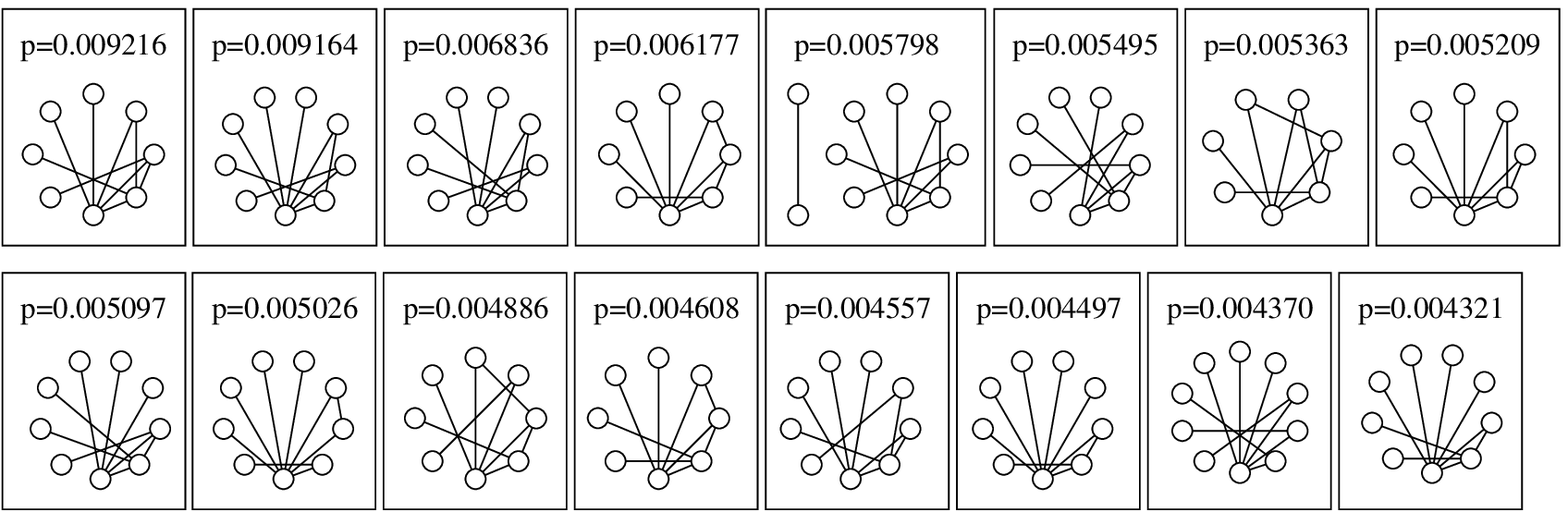}}
When the temperature is lowered to $T=0.3$, these are replaced by the
 following graphs :\\ \\
\centerline{\includegraphics[angle=0,width=0.9\columnwidth]{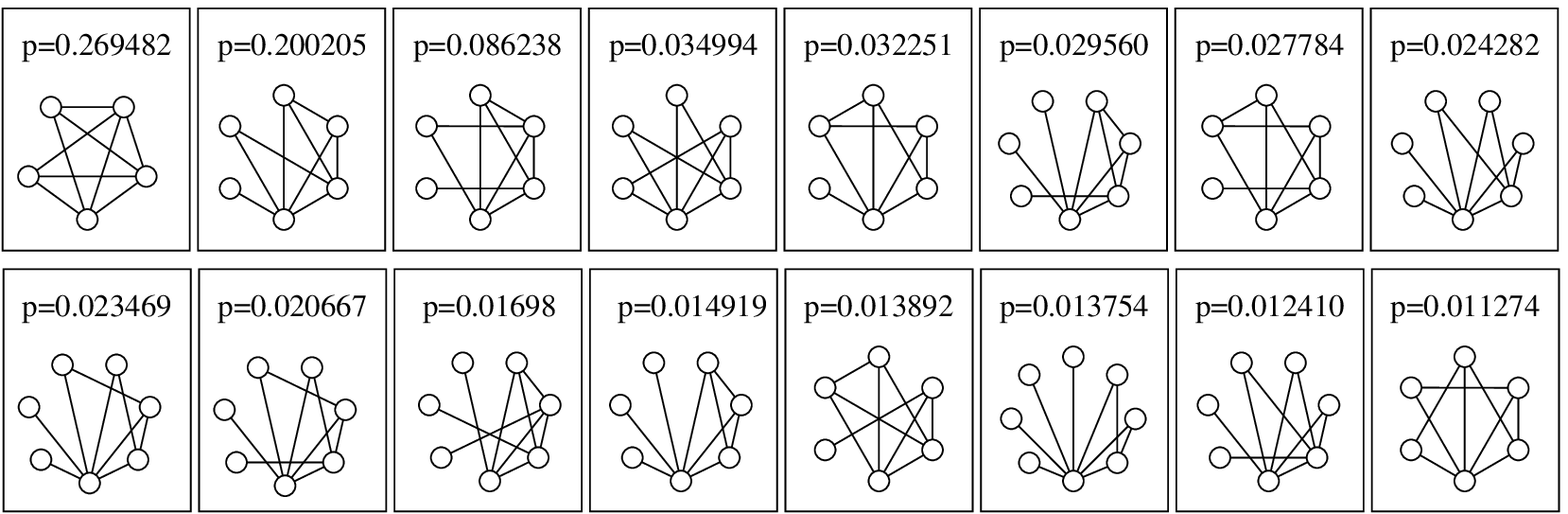}}

\section{Appendix B}

 For simplicity, we shall consider tree like 
clusters only and neglect the clusters with loops. Since the 
chances of a component containing a closed loop of edges goes as
$N^{-1}$ when $\kav<1$ and no giant connected component can be found in 
the system, this is a valid approximation 
in the thermodynamic limit \cite{N^-1}.
The number of possible trees of size $s$ in an undirected network can be 
estimated as follows. We pick
 a random realization of a tree sized $s$ 
(meaning $s$ edges and $s+1$ vertices), and we choose a vertex in it to be
the ``root'' of the tree. Starting from this root, we descend through 
all possible paths until we reach all the branches, and on the way we replace
 the undirected edges with directed ones pointing from the vertex closer
 to the root towards the vertex farther away from the root. This procedure
results in a directed tree, where each vertex (except the root) has one and
 only one incoming edge and $n\geq0$ outgoing edges. Then, another
 realization of a tree can be obtained from the present one by choosing a
 vertex, and moving the other end of the incoming edge from its original
place to a new vertex. Of course, this new vertex cannot be one of the 
``descendants'' of the selected vertex, since that way we would create
 a loop and
split the tree into two unconnected parts. 
Nevertheless, if $s$ is large enough, for
 the majority of the vertices this restriction eliminates only a negligible
 part of  the possible rewirings. Therefore we may estimate the number of 
possible new trees obtained from the rewiring of the incoming edge of 
a single vertex by $s$, and the total number of trees of size $s$ by $s^s$.

\end{document}